\documentclass[preprint2]{aastex6}
\usepackage{epstopdf}
\shorttitle{Radio emission from KISSR\,1219}
\shortauthors{Kharb et al.}
\begin{document}
\title{Double-Peaked Emission Lines due to A Radio Outflow in KISSR\,1219}
\author{P. Kharb}
\affil{National Centre for Radio Astrophysics - Tata Institute of Fundamental Research, Postbag 3, Ganeshkhind, Pune 411007, India}
\email{kharb@ncra.tifr.res.in}
\author{S. Subramanian}
\affil{Kavli Institute for Astronomy and Astrophysics, Peking University, 5 Yiheyuan Road, Haidian District, Beijing 100871, P. R. China}
\author{S. Vaddi}
\affil{National Centre for Radio Astrophysics - Tata Institute of Fundamental Research, Postbag 3, Ganeshkhind, Pune 411007, India}
\author{M. Das}
\affil{Indian Institute of Astrophysics, II Block, Koramangala, Bangalore 560034, India}
\author{Z. Paragi}
\affil{Joint Institute for VLBI in Europe, Postbus 2, 7990 AA Dwingeloo, the Netherlands}

\begin{abstract}
We present the results from {1.5 and 5~GHz} phase-referenced VLBA and 1.5~GHz {Karl G. Jansky Very Large Array (VLA)} observations of the Seyfert 2 galaxy KISSR\,1219, which exhibits double peaked emission lines in its optical spectrum. The VLA and VLBA data reveal a one-sided core-jet structure at roughly the same position angles, providing evidence of an AGN outflow. The absence of dual parsec-scale radio cores puts the binary black hole picture in doubt for the case of KISSR\,1219. The high brightness temperatures of the parsec-scale core and jet components ($>10^6$~K) are consistent with this interpretation. Doppler boosting with jet speeds of {$\gtrsim0.55c$ to $\gtrsim0.25c$}, going from parsec- to kpc-scales, at a jet inclination $\gtrsim50\degr$ can explain the jet one-sidedness in this Seyfert 2 galaxy. A blue-shifted broad emission line component in [O {\sc iii}] is also indicative of an outflow in the emission line gas at a velocity of $\sim350$~km~s$^{-1}$, while the [O {\sc i}] doublet lines suggest the presence of shock-heated gas. A detailed line ratio study using the MAPPINGS III code further suggests that a shock+precursor model can explain the line ionization data well. Overall, our data suggest that the radio outflow in KISSR\,1219 is pushing the emission line clouds, both ahead of the jet and in a lateral direction, giving rise to the double peak emission line spectra.
\end{abstract}

\section{Introduction}
\label{secintro}
Seyfert galaxies are the active galactic nuclei (AGN) subclass residing in spiral hosts. Their emission line spectra reveal the presence of prominent broad permitted and narrow permitted and forbidden emission lines. Those that reveal both broad and narrow lines are referred to as type 1, while those that reveal only narrow lines are termed type 2. Obscuration due to a dusty torus arising from orientation effects, is agreed to be largely responsible for the two primary types of Seyfert galaxies \citep{Antonucci93}. Seyferts have typically been classified as ``radio-quiet'' AGN \citep[$R\equiv S_\mathrm{5~GHz}/S_\mathrm{B~band}<10$;][]{Kellermann89} but frequently cross into the ``radio-loud'' class when their nuclear emission is properly extracted \citep{HoPeng01,Kharb14a}. A small fraction of Seyfert galaxies exhibit double-peaked emission lines in the optical/UV spectra \citep[e.g.,][]{Liu10}. The presence of double-peaked emission lines have been suggested to arise from gas in rotating disks \citep{Chen89,Eracleous03}, emission-line clouds being pushed away by bipolar outflows or complex narrow line region (NLR) kinematics \citep[e.g.,][]{Shen11}, and from binary broad- and narrow-line regions (BLR, NLR) around binary supermassive black holes \citep{Begelman80}.

As all luminous galaxies are believed to host supermassive black holes (BHs) in their centres \citep[M$_\mathrm{BH}\sim10^6-10^9$~M$_\sun$;][]{Kormendy13}, and galaxy mergers are an essential part of galaxy evolution, the presence of multiple black holes in galactic nuclei are expected \citep[e.g.,][]{Volonteri03}. However, so far, only $\sim$23 dual (BH separation $>$1 kpc)/binary (BH separation $\ll$1 kpc) AGN have been identified in the literature \citep{Deane14,Muller15}. The galaxy merger process which creates elliptical galaxies \citep[e.g.,][]{Steinmetz02}, results in gas infall, massive star formation and the formation of supermassive BH binaries. Spiral galaxies, on the other hand, are expected to undergo minor mergers (e.g., satellite accretion) which result in larger bulges but intact disks \citep{Aguerri01}. Binary BHs are therefore likely to be rare in spiral galaxies. This is consistent with the finding that almost all of the dual/binary AGN candidates reside in merger remnants or elliptical galaxies, with the exception of NGC\,3393 which resides in a spiral galaxy \citep{Fabbiano11}. 

Very Long Baseline Interferometry (VLBI) is currently the only technique by which binary BHs with projected separations of a few parsecs to few tens of parsecs, can be identified. Despite their radio weakness, several Seyfert galaxies have been observed with VLBI and parsec-scale cores and jets have been detected \citep[e.g.,][]{Nagar05,Kharb10a,Kharb14a,Mezcua14}. In this paper, we present results from VLBI observations of the Seyfert galaxy, KISSR\,1219, which exhibits double-peaked emission lines in its SDSS\footnote{Sloan Digital Sky Survey \citep{York00}.} spectrum.

The target galaxy for this study was selected from the Kitt Peak National Observatory (KPNO) International Spectroscopic Survey (KISS) which is an objective-prism survey of a large number of emission line galaxies \citep{Salzer00}. Spectroscopic observations of a sub-sample of 351 KISS objects were obtained by the 2.4m MDM telescope on Kitt Peak by \citet{Wegner03}; this forms the parent sample for our study. We examined the SDSS\footnote{The SDSS spectra are acquired through a fiber of diameter $3\arcsec$, or 2.2~kiloparsec at the distance of KISSR\,1219.} spectra of those galaxies which had been classified as either Seyfert type 1, type 2, or LINER\footnote{Low-Ionization Nuclear Emission-line Region galaxies} and identified six out of 65 galaxies ({\it i.e.}, 9\%) as having double peaks in their emission line spectra. Only three of these, viz., KISSR\,434, KISSR\,1219 and KISSR\,1494, had been detected in the FIRST\footnote{Faint Images of the Radio Sky at Twenty-Centimeters \citep{Becker95}} and NVSS\footnote{The NRAO VLA Sky Survey \citep{Condon98}} surveys at 1.4~GHz (resolution $\sim5.4\arcsec$ and $\sim45\arcsec$, respectively). We presented the results from a VLBI study of the brightest of these three, {\it viz.,} KISSR\,1494, in \citet{Kharb15b}. Here we present the results from phase-referenced multi-frequency VLBI observations of KISSR\,1219. New high resolution {Karl G. Jansky Very Large Array (VLA)} observations of KISSR\,1219 are also presented.

KISSR\,1219 (a.k.a. NGC\,4135) is a Seyfert type 2 galaxy residing in an SAB(s)bc type host: clear spiral arms and a weak bar are visible in its optical image. At its redshift of 0.037580, 1~milliarcsec (mas) corresponds to a linear extent of 0.729 parsec for H$_0$ = 73~km~s$^{-1}$~Mpc$^{-1}$, $\Omega_{mat}$ = 0.27, $\Omega_{vac}$ =  0.73. In this paper, spectral index, $\alpha$, is defined such that flux density at frequency $\nu$ is $S_\nu\propto\nu^\alpha$.

\section{Observations and Data Reduction}
\subsection{VLA Observations}
The VLA observations of KISSR\,1219 (Director's Discretionary Project 15A$-$468) were carried out on June 4, 2015 with the transition BnA$\rightarrow$A array-configuration at the L band (1.326~GHz). The full bandwidth was split into 64 channels with a bandwidth of 2 MHz each, and ten {SPWs (spectral windows)} ranging from 0.942 $-$ 1.966~GHz. The data integration time was 2 sec and the total on-source time was $\sim15$ mins. J1219+4829 and J1331+3030 were used as the phase and amplitude calibrators, respectively. We calibrated the data using the Common Astronomy Software Applications (CASA) package version 4.6.0. We followed the CASA tutorial on VLA wide-band wide-field imaging\footnote{https://casaguides.nrao.edu/index.php?title=EVLA\_Wide-Band\_Wide-Field\_Imaging:\_G55.7\_3.4}. Data from shadowed antennas and zero-amplitude data were flagged using {\tt FLAGDATA}. Before flagging the data affected by {RFI (radio frequency interference)}, we Hanning-smoothed the data using CASA task {\tt HANNINGSMOOTH}. We subsequently carried out automatic RFI excision using {\tt RFLAG} in {\tt FLAGDATA}, where we examined all the SPWs and scans individually. We eventually removed all data from {SPW 2 (centered at $\nu$ = 1.646 GHz)} and most of the data from {SPW 6 (centered at $\nu$ = 1.006 GHz)}, since it was heavily affected by RFI. Standard calibration tasks in CASA were used to calibrate the data. 

\begin{figure*}
\centering{\includegraphics[width=18cm]{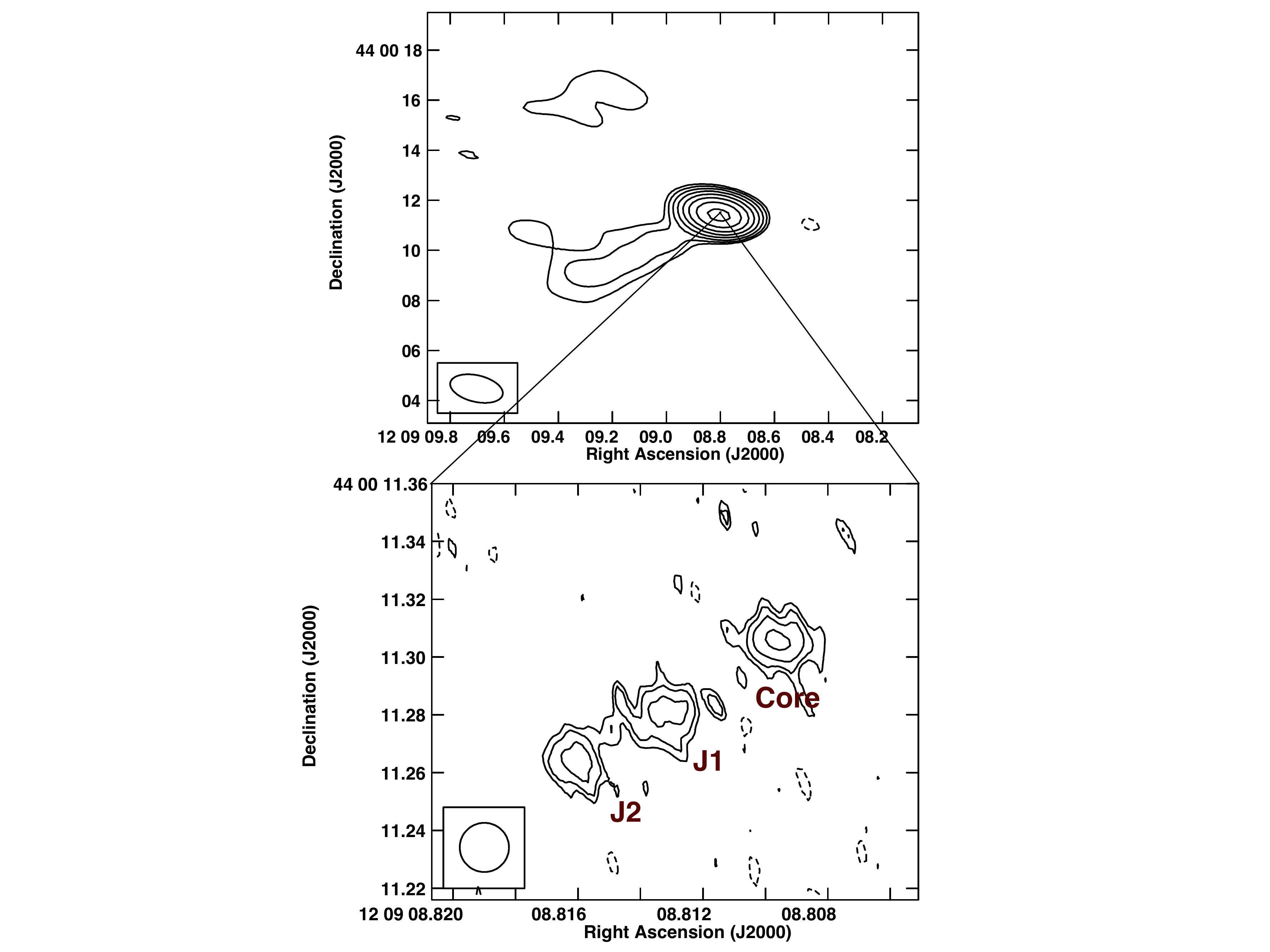}}
\caption{(Top) 1.5~GHz VLA and (Bottom) VLBA image of KISSR\,1219. The contours are in percentage of peak and increase in steps of $\sqrt 2$: the lowest contour level and peak surface brightness is (Top) $\pm$3.7\% and 7.13~mJy~beam$^{-1}$, (Bottom) $\pm$32\% and 0.26 mJy~beam$^{-1}$, respectively. The VLA beam-size is $2.14\arcsec\times1.06\arcsec$ at P.A.=77$\degr$, and the VLBA beam-size is $17~mas \times 17~mas$. }
\label{figvlbi}
\end{figure*}

We made several images using multi-scale clean, multi-scale-wide-field clean with W-projection, multi-scale-multi-frequency-synthesis, multi-scale-multi-frequency-wide-field clean, using {\tt imagermode =`csclean'} (Cotton-Schwab CLEAN) as well as {\tt `mosaic'}. The image made using multi-scale, multi-frequency-synthesis and {\tt imagermode=`mosaic'} had the lowest {\it rms} noise and highest source flux density. This is presented in the top panel of Figure~\ref{figvlbi}. Self-calibration did not work {for} this weak source.

\subsection{Phase-referenced VLBI}
{The VLBI observations were carried out in a phase-referencing mode with seven antennas\footnote{Hancock (HN), Owens Valley (OV) and Mauna Kea (MK) did not observe due to weather or motor issues.} of the Very Long Baseline Array (VLBA) at 1.5~GHz on February 06, 2015 (Project ID: BK191A), and eight antennas\footnote{Owens Valley and Mauna Kea did not observe due to weather or motor issues.} at 4.9~GHz on February 08, 2015 (Project ID: BK191B). The data were acquired with an aggregate bit rate of 2048 Mbits~sec$^{-1}$ (2 polarizations, 16 baseband channels, bandwidth 32~MHz, and a 2-bit sampling rate). 1218+444, which is 2.22$\degr$ away from the source and has an x,~y positional uncertainty of 0.16,~0.32 mas, was used as the phase reference calibrator. 4C\,39.25 and 1206+416 were used as the fringe-finder and phase-check calibrator, respectively. A nodding cycle of 5 mins (2 mins on the phase calibrator and 3 mins on the target) was used for the observations, interspersed by four 5 mins scans on 4C\,39.25 and one 3 mins scan on the phase-check calibrator. The experiment lasted a total of $\approx4.0$ hours at each frequency.} 

The data reduction was carried out using AIPS (Astronomical Image Processing System) following standard VLBA-specific tasks and procedures outlined in the AIPS cookbook\footnote{http://www.aips.nrao.edu/cook.html}. Los Alamos (LA), which was a stable antenna in the middle of the configuration, was used as the reference antenna for the calibration. The amplitude calibration was carried out using the procedure {\tt VLBACALA}, while the delay, rate, and phase calibration were carried out using the procedures {\tt VLBAPANG, VLBAMPCL} and {\tt VLBAFRGP}. The phase calibrator 1218+444 was iteratively imaged and self-calibrated on phase and phase+amplitude using AIPS tasks {\tt IMAGR} and {\tt CALIB}. The images were then used as models to determine the amplitude and phase gains for the antennas. These gains were applied to the target and the final images were made using {\tt IMAGR}. A round of data-flagging was carried out using the task {\tt CLIP} on the source {\tt SPLIT} file, prior to making the images. The radio component detected at 1.5~GHz was offset from the centre of the image by {$0.0\arcsec$, $-0.20\arcsec$} in right ascension and declination. We ran the task {\tt UVFIX} on the source {\tt SPLIT} file to shift the source to the centre before producing the final image. We used the elliptical Gaussian-fitting task {\tt JMFIT} to obtain the component flux densities. 

As the source was weak at 1.5~GHz, we specified three boxes (using parameter {\tt CLBOX}) in the task {\tt IMAGR} to run {\tt CLEAN}. The choice of these three components was made to produce an image with low and uniform residual noise. {Self-calibration could not be carried out for this weak source.} The final naturally-weighted image made using {{\tt ROBUST=0}} and a circular beam of 17~{\it mas}, is presented in the bottom panel of Figure~\ref{figvlbi}. 
\begin{figure*}
\centerline{\includegraphics[width=17cm,trim=0 350 0 35]{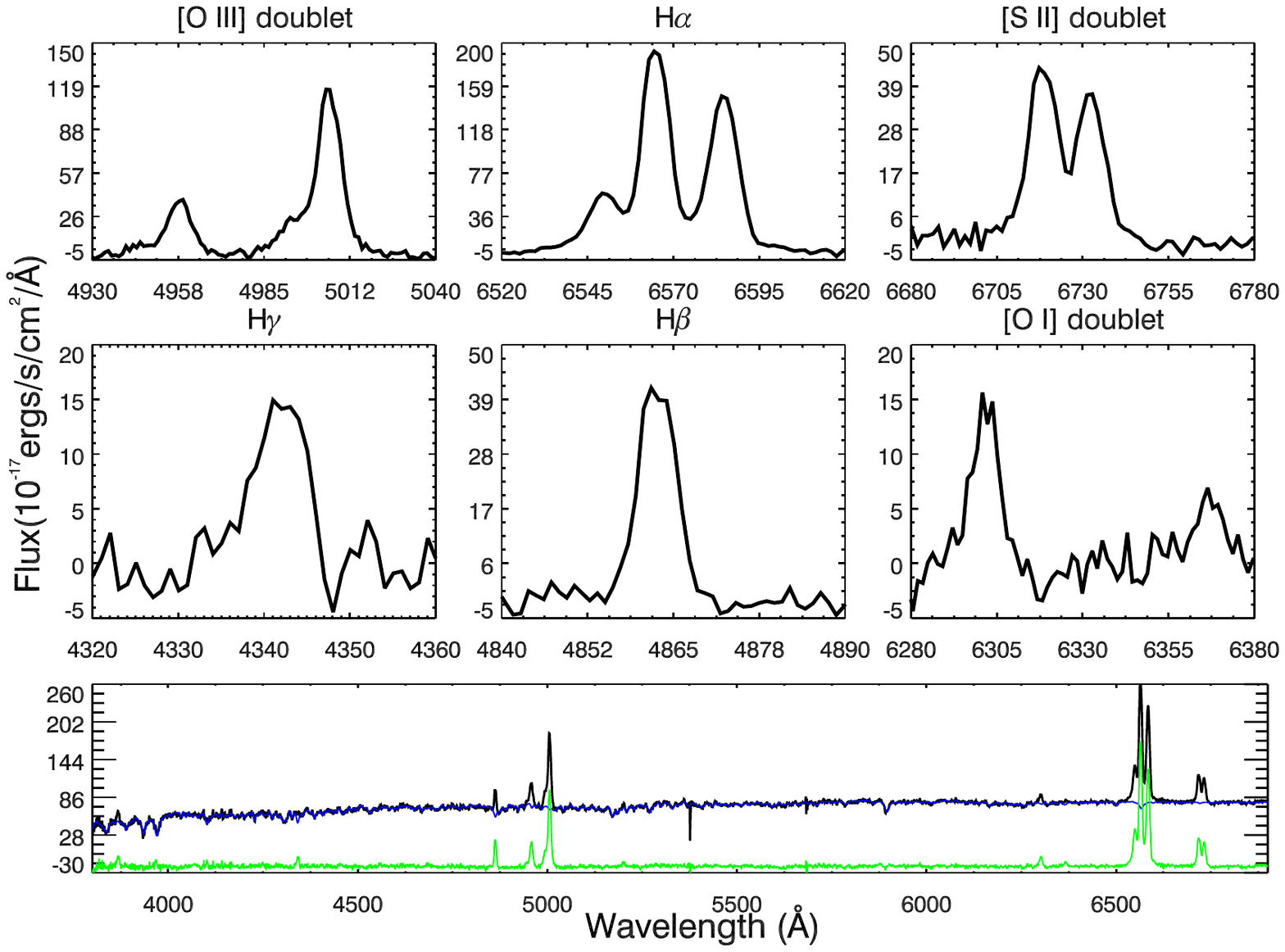}}
\caption{SDSS spectrum of KISSR\,1219 showing the de-reddened double-peaked emission lines. The bottom panel displays the entire spectrum along with the fitted stellar continuum in blue. The residue of the stellar continuum fit or the pure emission line spectrum is also shown in green. For better visualisation, the pure emission line spectrum is shifted by $-20$ in y-axis.}
\label{figspectra1}
\end{figure*}

\subsection{Emission-Line Fitting}
The observed SDSS optical spectrum was corrected for reddening using the E(B$-$V) value from \citet{Schlegel98}. The {\tt pPXF} (Penalized Pixel-Fitting stellar kinematics extraction) code by \citet{Cappellari04} was used to obtain the best-fit model of the underlying stellar population, in order to isolate the AGN emission lines. Emission lines in the de-reddened spectrum were masked and the underlying absorption spectrum was modelled as a combination of single stellar population templates with MILES \citep{Vazdekis10}; these templates are available for a large range of metallicity (M/H $\sim-2.32$ to +0.22) and age (63 Myr to 17 Gyr). 

{\tt pPXF} provides the best-fit template and the stellar velocity dispersion of the underlying stellar population ($\sigma_\star$), which was 142$\pm$6.9 km~s$^{-1}$ for KISSR\,1219. Although the major contribution to the observed spectrum is the underlying stellar population ($\sim$80\%), there could be other contributors like power law component from AGN and Fe lines ($\sim$20\%). While estimating the best-fit to the underlying population, {\tt pPXF} fits a polynomial along with the optimal template to account for the contribution from the power law continuum. The Fe lines are too weak to affect further analysis. The output of the analysis is shown in the lower panel of Figure~\ref{figspectra1}. The reddening-corrected spectra is shown in black and the best-fit model is over-plotted in blue. The best-fit model is subtracted from the de-reddened  spectrum to obtain the pure emission line spectrum, shown in the lower panel in green. The prominent emission lines  in the pure emission line spectrum such as, [S {\sc ii}] $\lambda$$\lambda$6717, 6731 doublet, [N {\sc ii}] $\lambda$$\lambda$6548, 6584 doublet, H$\alpha$, [O {\sc i}]$\lambda$$\lambda$6300, 6364 doublet, [O {\sc iii}]$\lambda$$\lambda$5007, 4959 doublet, H$\beta$ and H$\gamma$ are shown in the middle and upper panels of Figure~\ref{figspectra1}. Signatures of double peaks are seen in all these emission lines.
  
\begin{figure*}
\centering{
\includegraphics[width=17cm,trim=0 450 0 160]{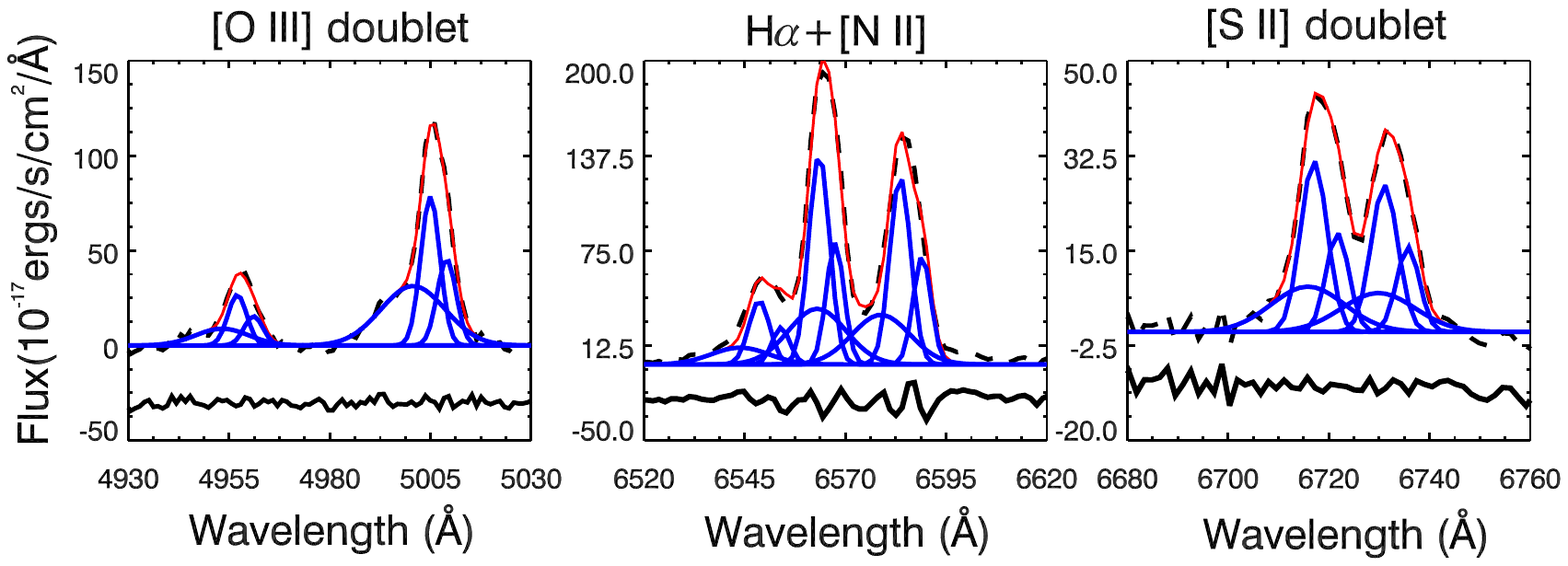}
\includegraphics[width=17cm,trim=0 430 0 150]{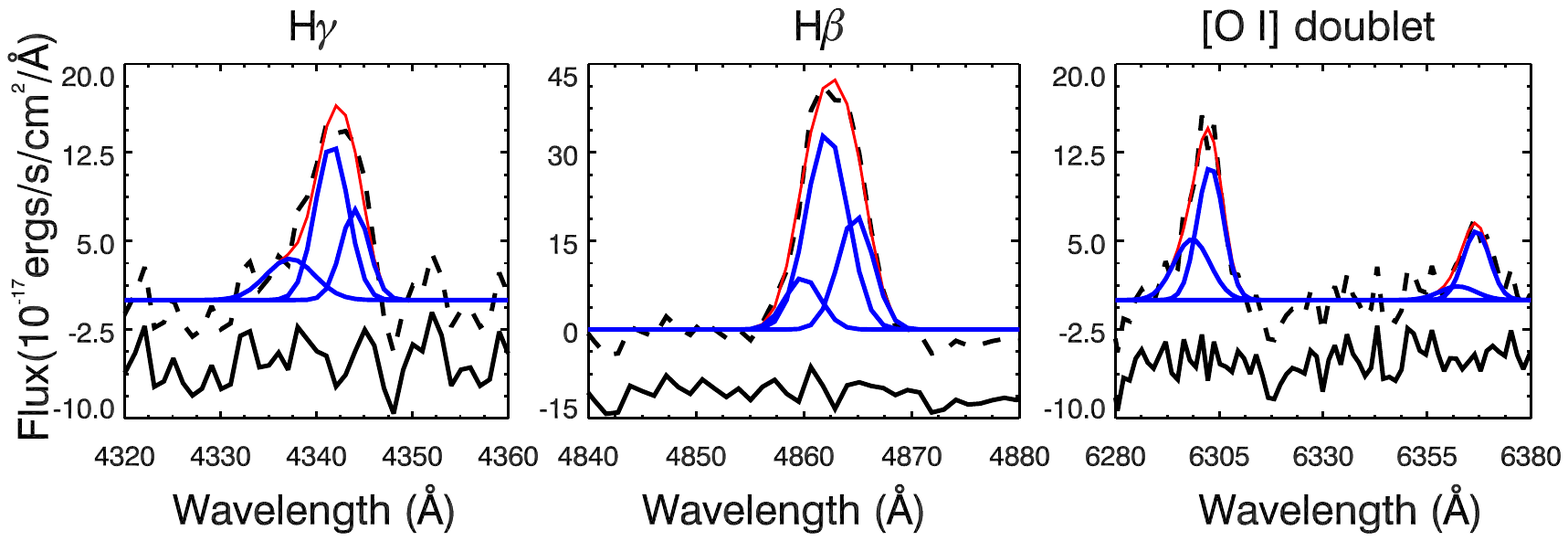}
\caption{(Top row) SDSS spectrum of KISSR\,1219 showing the dereddened double-peaked H$\gamma$, H$\beta$ and  [O {\sc i}] in black (dashed lines). (Bottom row) SDSS spectrum showing the dereddened double-peaked [O {\sc iii}], H$\alpha$ + [N {\sc ii}] and [S {\sc ii}] lines in black (dashed lines). The individual Gaussian components are shown in blue and the total fit to the lines in red. In all the panels, residuals of the fit are shown in black solid line.}
\label{figspectra2}}
\end{figure*}

In order to measure the line parameters, the pure emission line spectrum is analysed by modelling the profiles as Gaussians. In general, the [S {\sc ii}] doublet lines which are well separated are considered as good representation of the shape of the [N {\sc ii}] and H$\alpha$ narrow lines \citep{Filippenko88,Greene04}. Hence, we first modelled the [S {\sc ii}] lines to obtain a satisfactory fit to the line-shape and used that model as a template for other narrow lines. The two [S {\sc ii}] lines are assumed to have equal width (in velocity space) and are separated by their laboratory wavelengths. We also tried to fit each [S {\sc ii}] line with double and triple component Gaussian models. When multiple Gaussian components are included, the corresponding components of each line are assumed to have equal width (in velocity space) and are separated by their laboratory wavelengths. Also, the relative intensities of different components of each line are held fixed. The best-fit to each [S {\sc ii}] line required three Gaussian components, as shown in the lower right panel of Figure~\ref{figspectra2}. The two narrow components in each [S {\sc ii}] line represent the two lines corresponding to the double peak feature. The third component represents the extended wing, to which we do not attribute any physical significance. This component was essential to fully describe the [S {\sc ii}] line-shape and its inclusion is statistically justified by an improvement in the reduced $\chi^2$ value by $\sim$20\%. 

The [S {\sc ii}] model was finally used as a template to fit the narrow H$\alpha$ and [N {\sc ii}] doublet lines. The widths of the different components of the [N {\sc ii}] doublet and H$\alpha$ narrow lines were assumed to be the same as that of the corresponding components of the [S {\sc ii}] lines. The separation between the centroids of the [N {\sc ii}] narrow components were held fixed and the flux of [N {\sc ii}] $\lambda$6583 to [N {\sc ii}] $\lambda$6548 was fixed at the theoretical value of 2.96. As there are three components for each [S {\sc ii}]  line, the profiles of H$\alpha$ and [N {\sc ii}] doublet narrow lines were strictly scaled from [S {\sc ii}]. The best-fit profile and individual components to the H$\alpha$ and [N {\sc ii}] narrow lines are shown in the lower-middle panel of Figure~\ref{figspectra2}.

The H$\beta$ and H$\gamma$ line profiles were also modelled using the [S {\sc ii}] template, but now the width of the third Gaussian component corresponding to the extended wing, was kept as a free parameter. As the fit improved by 18\% in reduced $\chi^2$ compared to the model where we kept the width fixed, and as we do not attribute any physical significance to the third Gaussian component, the choice to keep the width as a free parameter was justified. The best-fit profile and individual components to the H$\beta$ and H$\gamma$ lines are shown in upper-middle and upper-left panels of Figure~\ref{figspectra2}, respectively. 

As the [O {\sc iii}] profile does not typically match with that of [S {\sc ii}] \citep{Greene05}, we fit the [O {\sc iii}]  and [O {\sc i}] independently. We initially used two Gaussian component model and a third additional Gaussian component was included if it improved the corresponding fit by {$>20\%$}. As shown in the lower-left and upper-right panels of Figure~\ref{figspectra2}, each line of the [O {\sc iii}] doublet is modelled using three Gaussian components  and the [O {\sc i}] doublet lines using two Gaussian components. In the case of the [O {\sc iii}] doublet lines, the third component was significantly blue-shifted and broad. Hence, these lines may be representing an outflow feature.

IDL programs which use the {\tt MPFIT} function for non-linear least-square optimisation were used to fit the emission line profiles with Gaussian functions and to obtain the best-fit parameters. The fit parameters corresponding to each line are presented in Table~\ref{tabprop}. The errors associated with each parameter are those provided by {\tt MPFIT}. 

\begin{figure*}
\centering{
\includegraphics[width=9cm,trim=0 470 50 180]{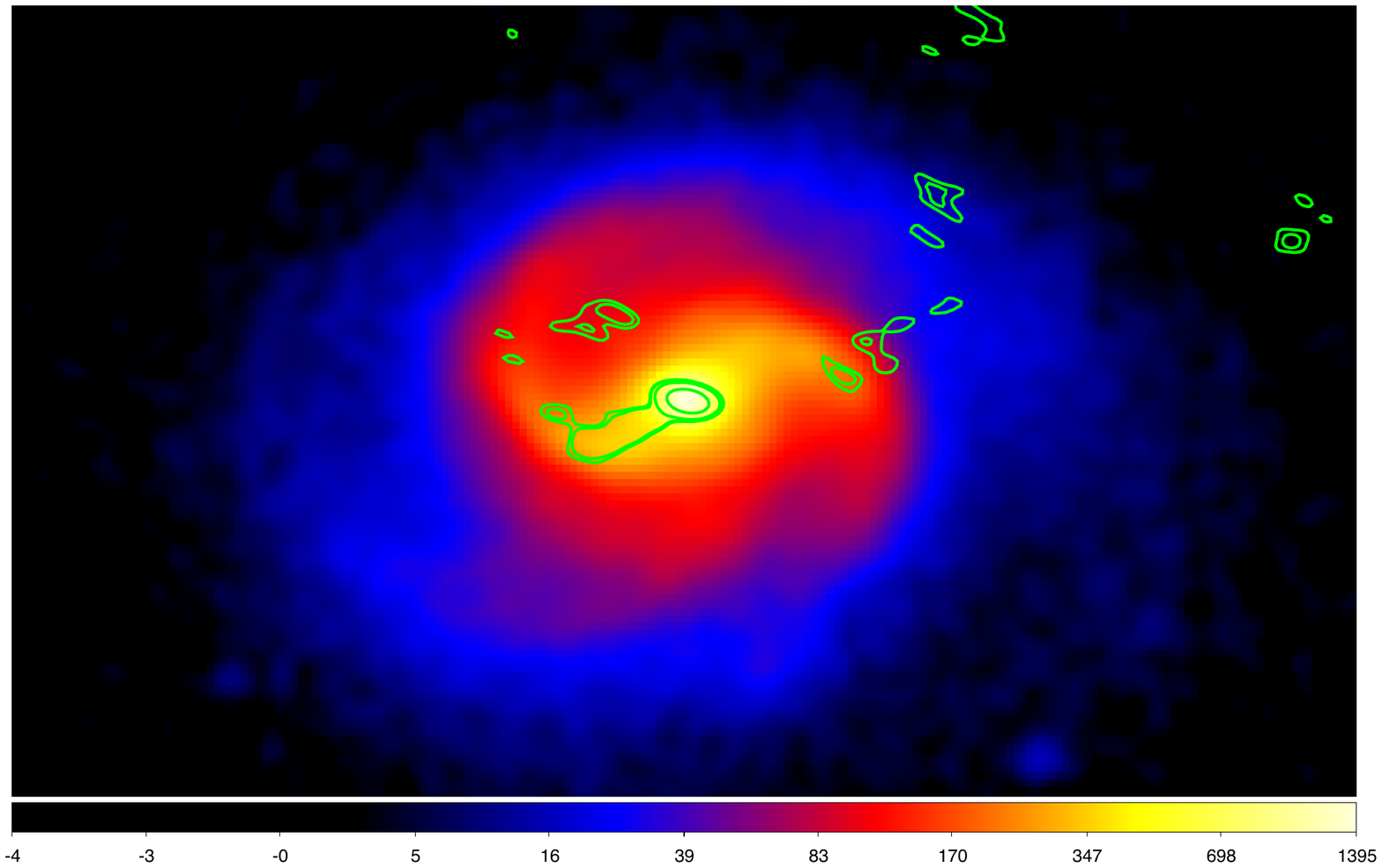}
\includegraphics[width=9cm,trim=0 230 60 0]{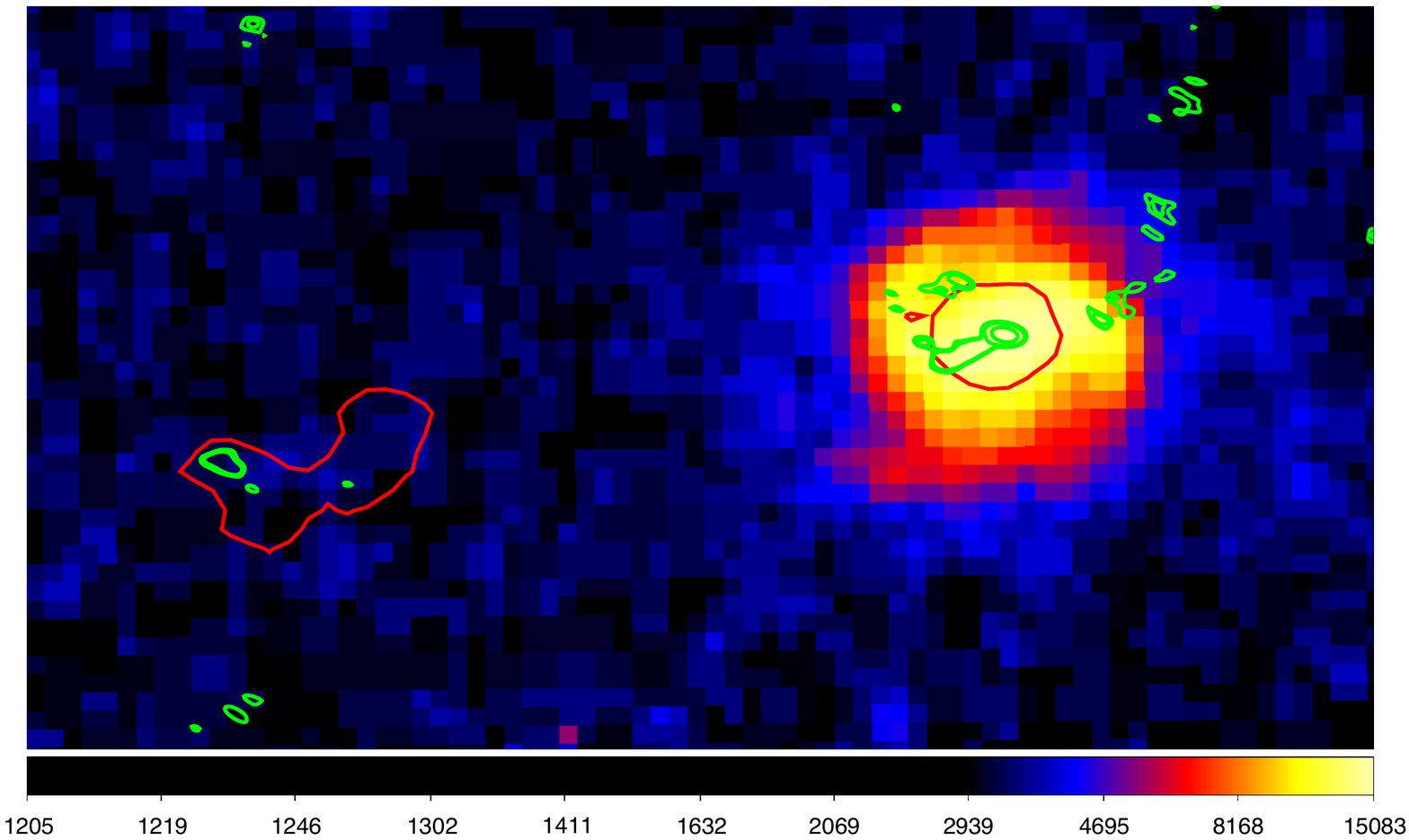}}
\caption{Radio-optical overlay of KISSR\,1219. (Top) Green contours representing the new 1.5~GHz VLA emission are superimposed on the SDSS image of the host galaxy. (Bottom) Red contours representing the 1.4~GHz FIRST emission and green contours the new 1.5~GHz VLA emission are superimposed on the optical image from DSS. {Some emission was detected coincident with the diffuse component seen in the FIRST image, about $\approx63\arcsec$ (46~kpc) to the east of KISSR\,1219, at a P.A. similar to the radio jet. This is likely to be an unconnected radio source (see Section~\ref{seckpcjet}).}}
\label{figradopt}
\end{figure*}

\section{Results}
\label{secresults}
The 1.5~GHz VLA image reveals a one-sided jet at a position angle (P.A.) of $\sim120\degr$, which is along the same direction as one of the spiral arms of the optical host galaxy. (Note that the P.A. is measured from North through East on the sky, with North being at 0$\degr$.) However, the jet differs in structure from the spiral arm in that it is straight and does not follow the spiral arm curve precisely. Its total extent is around $7\arcsec$ ($\sim$5 kpc). Faint diffuse emission, previously observed in the 1.4~GHz FIRST image, was also detected $\approx63$ arcseconds (46~kpc) away (see Figure~\ref{figradopt}). This is discussed {in Section~\ref{seckpcjet}}.

The VLBA image at 1.5~GHz shows a parsec-scale core-jet-like structure. Three distinct components are observed in the 1.5~GHz image (Figure~\ref{figvlbi}, bottom panel). We identify the brightest of these components as the radio ``core''. The total extent of the core-jet structure is around 100 mas ($\sim$70 parsec). Its P.A. is almost identical to the P.A. of the VLA jet. The core-jet structure is not detected at 5~GHz, implying an overall steep radio spectrum. {The VLBA picks up nearly $70$\% of the total flux detected by the VLA. Therefore, nearly 30\% of the radio emission could be on intermediate scales between the VLA and the VLBA.} The absence of dual compact radio cores puts the binary BH picture in serious doubt for the case of KISSR\,1219.

The SDSS spectrum shows a clear blue-shifted broad component in the [O {\sc iii}] emission line, which is indicative of an outflow. The presence of the [O {\sc i}] doublet lines is also indicative of shock-heated gas. A connection between the radio outflow and emission line gas outflow is therefore implicated in KISSR\,1219. {We discuss this in more detail in Section~\ref{secdisc}.}

\subsection{The Parsec-scale Radio Structure}
\label{secparsec}
The 1.5~GHz VLBA radio core position of KISSR\,1219 is at RA 12h 09m 08.80956s, DEC 44$\degr$ 00$\arcmin$ 11.3057$\arcsec$. {The peak intensity of the 1.5~GHz radio core is $\sim$0.26~mJy~beam$^{-1}$; the total flux density of the entire core-jet structure is $\sim$1.58~mJy. While KISSR\,1219 was not detected at 5~GHz with the VLBA, the phase calibrator 1218+444 was detected. The final {\it rms} noise at 1.5 and 5~GHz is $\sim40$~$\mu$Jy~beam$^{-1}$ and $\sim30$~$\mu$Jy~beam$^{-1}$, respectively.}

Using the total flux densities of the core and two jet components J1 and J2, the brightness temperature ($T_B$) at 1.5~GHz \citep[see the relation in][]{Ulvestad05} turn out to be $\sim6\times10^6$~K, $\sim3\times10^6$~K, and $\sim4\times10^6$~K, respectively, {for the unresolved components}. These relatively high brightness temperatures support non-thermal AGN-related emission. While VLBI has indeed detected individual radio supernovae and supernova remnants in starburst galaxies with similarly high brightness temperatures \citep[$\ge10^6$~K,][]{Lonsdale06,Perez09}, the close alignment of the parsec- and kpc-scale radio jets in KISSR\,1219, support the AGN picture instead.

{As no core-jet emission}
was detected in the 5~GHz image, we infer the $1.5-5$~GHz spectral indices to be steeper than $-1.2, -1.1$ and $-1.0$ for the core, J1 and J2, respectively, assuming {three} times the {\it rms} noise in the 5~GHz image as an upper limit to the flux density of these components. The errors in the spectral index values are $\approx20\%$. It is known that VLBI ``cores'' in some Seyferts do not exhibit flat or inverted spectra and can have steep spectra instead \citep[e.g.,][]{Roy00,Orienti10,Kharb10a,Bontempi12,Panessa13}. This is unlike the cores in powerful radio sources, which are suggested to be the unresolved bases of relativistic jets. 

We remade the VLBA images of KISSR\,1219 using different weighting schemes by varying the ROBUST parameter, in order to check for the compactness of the core. We found that the core, or rather a compact portion of it, remained intact and centrally concentrated in all the images, even as the jet components (J1 and J2) got largely resolved out in ROBUST $-5$ (pure uniform weighting) images. This result supports the suggestion that the brightest feature in the bottom panel of Figure~\ref{figvlbi}, is indeed the radio core, albeit with partially-resolved jet emission in close proximity, which makes the overall core spectrum steep. 

\subsection{Kpc-scale Jet Energetics}
\label{seckpcjet}
{The peak intensity of the 1.5~GHz VLA radio core is $\sim$3.35~mJy~beam$^{-1}$; the total flux density of the entire core-jet structure is $\sim$2.33~mJy. The final {\it rms} noise in the 1.5~GHz VLA image is $\sim100$~$\mu$Jy~beam$^{-1}$.}
Assuming equipartition of energy between relativistic particles and the magnetic field \citep{Burbidge59}, we obtained the minimum {pressure, and the} particle energy (electrons and protons) at minimum pressure using the relations in \citet{OdeaOwen87}. The total radio luminosity was estimated assuming that the radio spectrum extends from ($\nu_l=$) 10~MHz to ($\nu_u=$) 100~GHz with a spectral index of $\alpha=-1$. It was assumed that the ratio of the relativistic proton to relativistic electron energy was unity. Table~\ref{tabequip} lists equipartition estimates for plasma filling factors of unity and 0.5 \citep[e.g.,][]{Blustin09}. The total energy in particles and fields, $E_{tot}$, is estimated as $E_{tot}=1.25\times E_{min}$, while the total energy density, $U_{tot}$, in erg~cm$^{-3}$ is $=E_{tot}(\phi V)^{-1}$. The minimum energy magnetic fields are of the order of $\sim$1~mG on the VLBA scales and $\sim$30~$\mu$G on the VLA scales, while the total energy $E_{tot}$ is $\approx4\times10^{53}$~ergs and $\approx7\times10^{55}$~ergs, respectively. The lifetime of electrons in the radio component undergoing both synchrotron radiative and inverse Compton losses on CMB photons was estimated using the relation in \citet{vanderlaan69}. These are of the order of 10,000 yr on the VLBA scales and $\sim$4~Myr on the VLA scales.

If we assume that the kiloparsec-scale emission is emission from a broad AGN outflow in KISSR\,1219, then we can estimate the time-averaged kinetic power ($Q$) of this outflow following the relations for radio-powerful sources listed by \citet{Punsly11} \citep[see also][]{Willott99}: we derive $F_{151}$ using the 1.4~GHz flux density from FIRST and a jet/lobe spectral index of $-0.8$, and obtain a kinetic power of $Q\approx2.5\times10^{41}$~erg~s$^{-1}$. This $Q$ value is typical of outflows in low-luminosity AGN \citep[e.g.,][]{Mezcua14}.

We did not see any connection between the VLA jet and the diffuse emission $\approx63$ arcseconds (46~kpc) to the east of KISSR\,1219, and at a P.A. similar to the radio jet (Figure~\ref{figradopt}). Instead we find a weak DSS optical source but a strong WISE infrared source (in bands 1 and 2 at 3.4$\mu$m and 4.6$\mu$m, respectively) in the center of the diffuse emission. We therefore conclude that this diffuse emission is unrelated to KISSR\,1219. It comes from a possibly higher redshift radio galaxy.

\subsection{Jet One-sidedness}
We can rule out the radio structure on kpc-scales to be emission from the spiral arms of the galaxy, on the basis of its one-sidedness. As the galaxy is nearly face-on, it is difficult to explain why the counter spiral arm is not emitting radio waves. The straight VLA radio feature which is exactly coincident with the VLBA core-jet structure is consistent with it being a synchrotron jet from the AGN in KISSR\,1219. 

In order to test if the one-sided VLA jet in KISSR\,1219 could be a result of relativistic beaming effects, we estimated the jet-to-counterjet surface brightness ratio ($R_J$) at a distance of 6$\arcsec$ (=4.37 kpc) from either side of the core ($\sim$0.46~mJy~beam$^{-1}$ on the jet side versus $\sim$0.19~mJy~beam$^{-1}$ on the counterjet side). For the jet structural parameter of $p=3.0$ \citep[continuous jet with $\alpha=-1$;][]{Urry95}, $R_J$ is 2.5. We can estimate a similar $R_J$ on VLBI scales; $R_J = 8.5$ for the jet component (J1) at a distance of $\sim40$~mas (=30 pc) from the radio core. Assuming that the torus half-opening angle is $\sim50\degr$ \citep[e.g.,][]{Simpson96}, and therefore the orientation of KISSR\,1219 should be greater than $\sim50\degr$ for it to be classified as a Seyfert 2, we can derive lower limits on the jet speeds on both parsec- and kpc-scales, in order to produce the observed $R_J$ values. The lower limits of jet speeds are $\sim0.55c$ and $\sim0.25c$ on the VLBA and VLA scales, respectively. As these are viable values of jet speeds in Seyfert galaxies \citep{Ulvestad99}, we cannot rule out Doppler boosting (and dimming) effects being responsible for the one-sided jet structure in KISSR\,1219.

If the missing counterjet emission on the other hand, is a result of free-free absorption, the required electron densities of the ionized gas on both parsec- and kpc-scales can be estimated, using the relations, $EM = 3.05\times10^6~\tau~T^{1.35}~\nu^{2.1}$, and $n_e=\sqrt{EM/l}$ \citep{Mezger67}. Here $EM$ is the emission measure in pc cm$^{-6}$, $\tau$ is the optical depth at frequency $\nu$ in GHz, $T$ the gas temperature in units of $10^4$~K, $n_e$ the electron density in cm$^{-3}$ and $l$ the path length in parsecs. In order to account for the observed jet-to-counterjet surface brightness ratios of 8.5 on parsec-scales and 2.5 on kpc-scales, the optical depth at 1.5~GHz should be at least $\sim$2 and $\sim$1, respectively.\footnote{{using $\exp(-\tau)=1/R_J$, for example see \citet{Ulvestad99}}} For a gas temperature of $10^4$~K and a path length of 1~pc and 100~pc for the VLBA and VLA jets, respectively, $EM$ of $\approx1.4\times 10^7$~pc~cm$^{-6}$ and $7.1\times 10^6$~pc~cm$^{-6}$, and $n_e$ of $\approx$4000~cm$^{-3}$ and 250~cm$^{-3}$ are required for free-free absorption, on the VLBA and VLA scales, respectively. Such ionized densities can indeed come from gas clouds in the narrow line region (see also Section~\ref{secnlr} ahead). However the volume filling factor of NLR clouds is small $-$ of the order of $10^{-4}$ \citep[e.g.,][]{Alexander99}, {making them unlikely candidates for absorbers, especially on kpc-scales where the jet width becomes larger ($>1$~kpc).} Ionized gas in giant HII regions with $n_e\sim100-1000$~cm$^{-3}$ and lifetimes $\sim10^7$ yr could in principle also be the candidate media for free-free absorption \citep[][]{Clemens10}. 
{The volume filling factor of this gas is also small, $\ge0.2$ \citep[e.g.,][]{Walterbos94}. We conclude that Doppler boosting/dimming effects are a more favourable explanation for the one-sided jet structure observed in KISSR\,1219.}

\subsection{Black Hole Mass \& Accretion Rate} \label{secbh}
{The bulge stellar velocity dispersion as derived by our line-fitting is $\sigma_\star=142.0\pm6.9$~km~s$^{-1}$. Following the $M_{BH}-\sigma_\star$ relation for late-type galaxies by \citet{McConnell13}:
$\mathrm{log(\frac{M_{BH}}{M_\sun})=8.07+5.06~log\,(\frac{\sigma_\star}{200~km~s^{-1}})}$, 
the mass of the black hole in KISSR\,1219 turns out to be $2.1\pm1.5\times10^7~\mathrm{M_{\sun}}$. The Eddington luminosity ($\mathrm{L_{Edd}\equiv1.25\times10^{38}~M_{BH}/M_\sun}$) is $\approx2.6\times10^{45}$~erg~s$^{-1}$. The bolometric luminosity estimated using the [O {\sc iii}]~$\lambda$5007 line flux  is $\sim6.19\times10^{43}$~erg~s$^{-1}$. The Eddington accretion rate ($\mathrm{\equiv L_{bol}/L_{Edd}}$) is $\sim0.02$, which is typical of low luminosity Seyfert galaxies \citep{Ho08}. 
The mass accretion rate $\mathrm{\dot{M}_{acc} = L_{bol}/\eta c^2}$ for this galaxy is $\sim0.01$~M$_{\odot}$~yr$^{-1}$, assuming a mass-to-energy conversion efficiency factor $\eta=0.1$.

\begin{figure*}
\includegraphics[width=9cm]{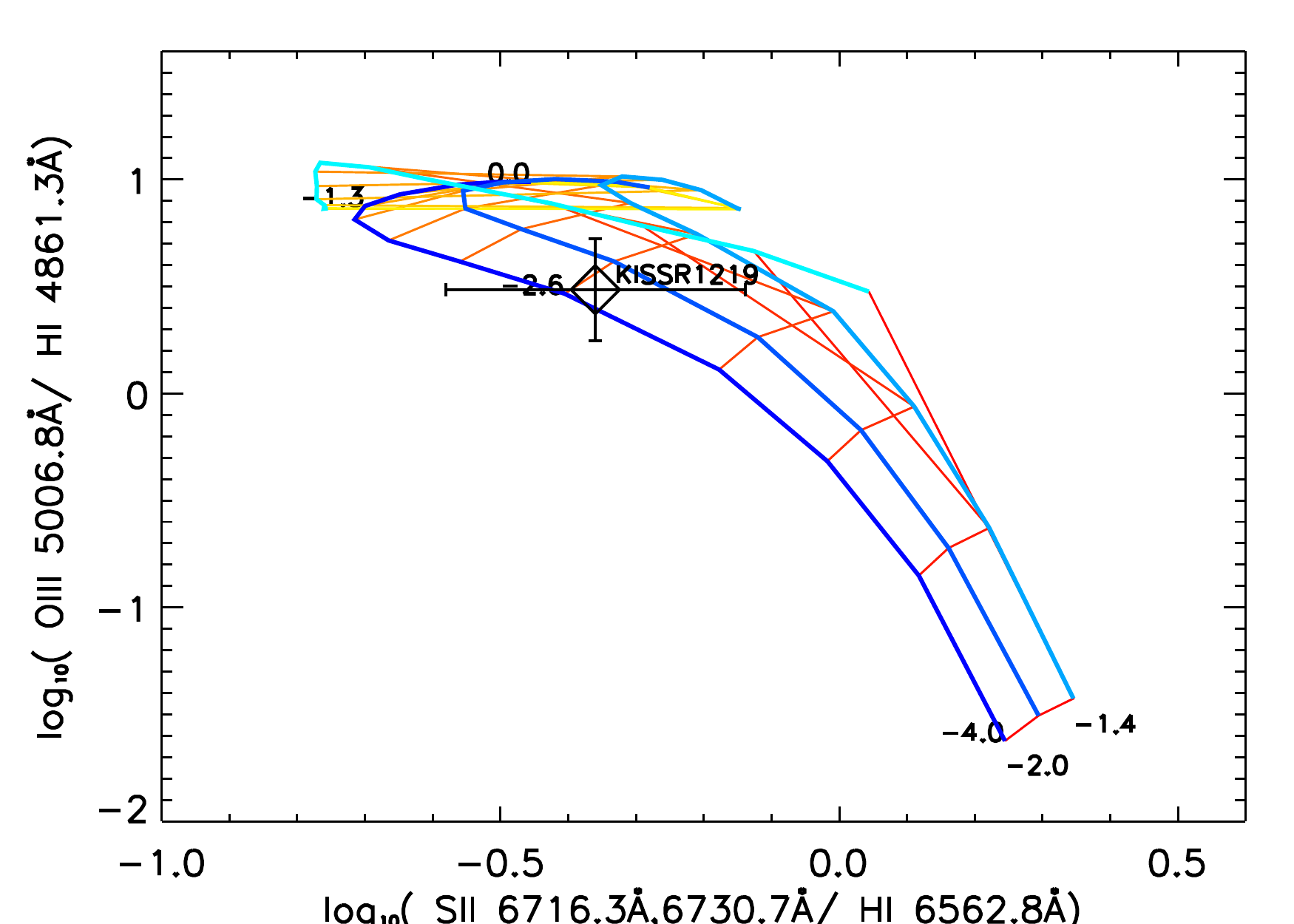}
\includegraphics[width=9cm]{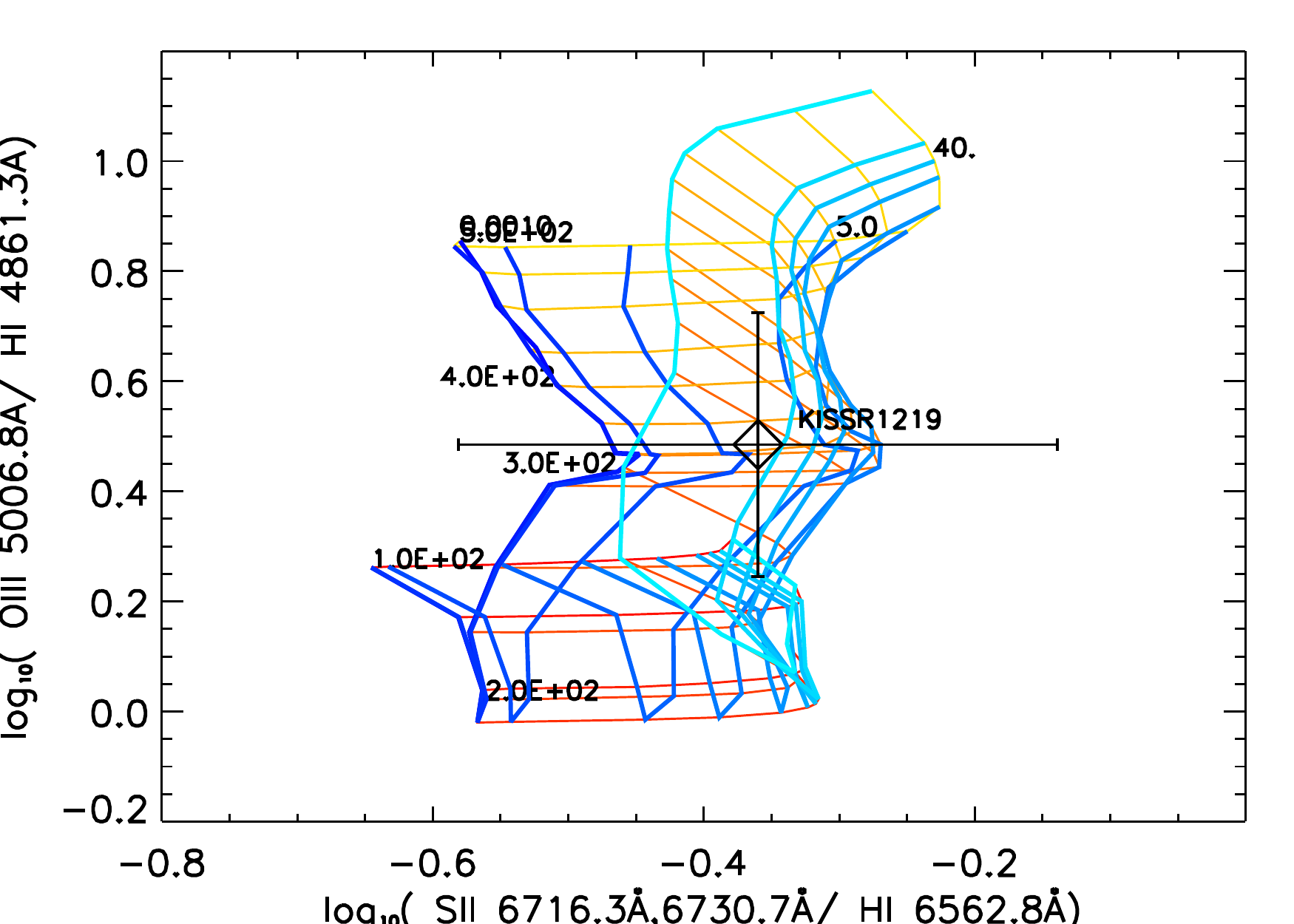}
\caption{[S {\sc ii}] $\lambda$6716, 6731/H$\alpha$ vs [O {\sc iii}] $\lambda$5007/H$\beta$ diagnostic diagram for a density of 100~cm$^{-3}$ and solar abundance. (Left) AGN Photoionization model grid of varying ionization parameters ($-4<$ Log U $<0$) and power-law indices ($-2< \alpha <-1.2$). {Lines of constant $\alpha$ are shaded in blue and lines of constant U are red-yellow shaded.} (Right) Shock+precursor model grid for varying magnetic field parameters {($0.001 <$ B $< 100$ $\mu$G~cm$^{3/2}$) and shock velocities ($100 < \,v \, < 500$~km~s$^{-1}$).  Lines of constant magnetic field are blue shaded and lines of constant shock velocities are red-yellow shaded.}}
\label{fig:sii_ha_oiii_hb}
\end{figure*}

\begin{figure*}
\includegraphics[width=9cm]{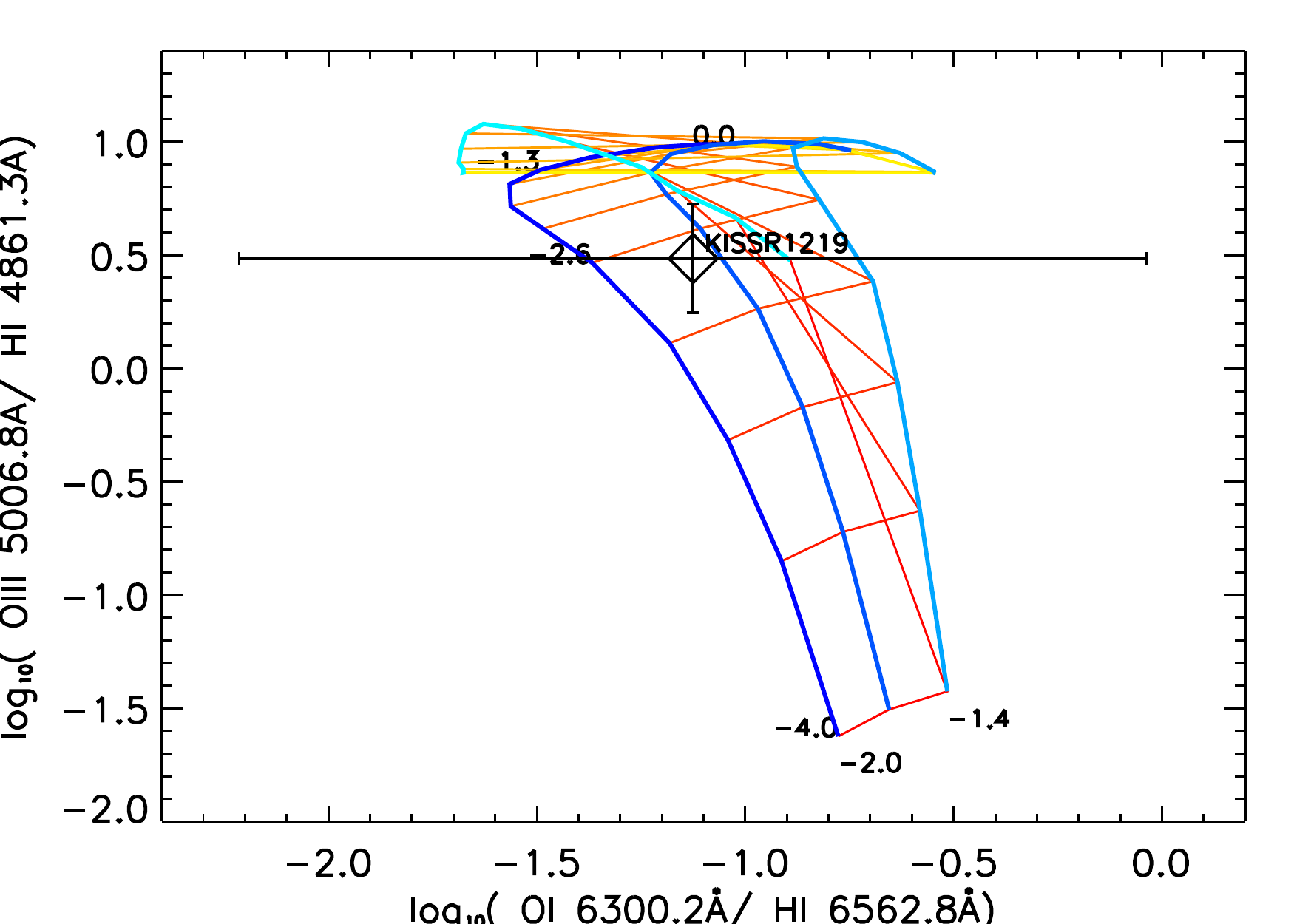}
\includegraphics[width=9cm]{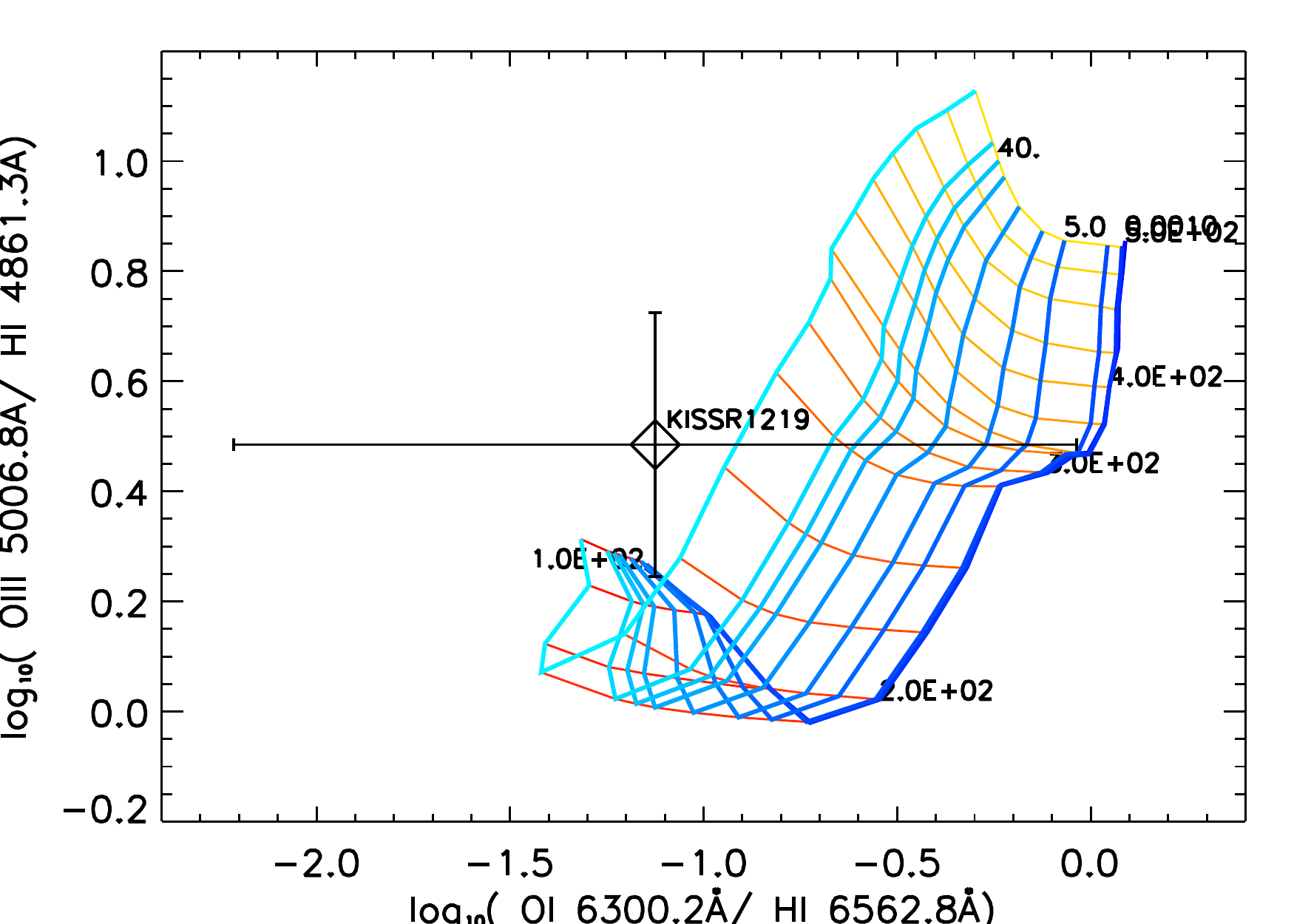}
\caption{[O {\sc i}] $\lambda$6300/H$\alpha$ vs [OIII] $\lambda$5007/ H$\beta$.  Details same as in Figure ~\ref{fig:sii_ha_oiii_hb}.} 
\label{fig:oi_ha_oiii_ha}
\end{figure*}

\begin{figure*}
\includegraphics[width=9cm]{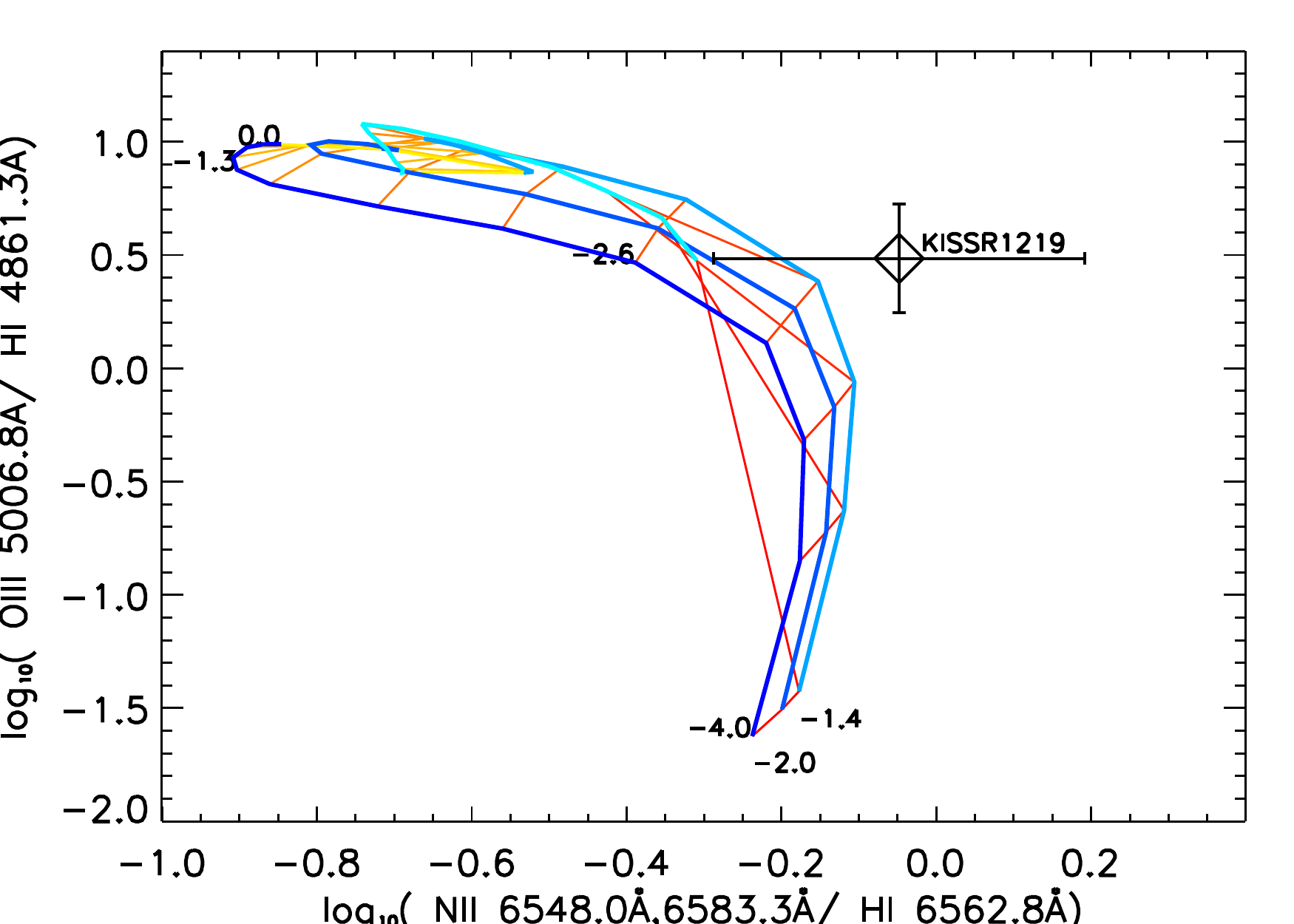}
\includegraphics[width=9cm]{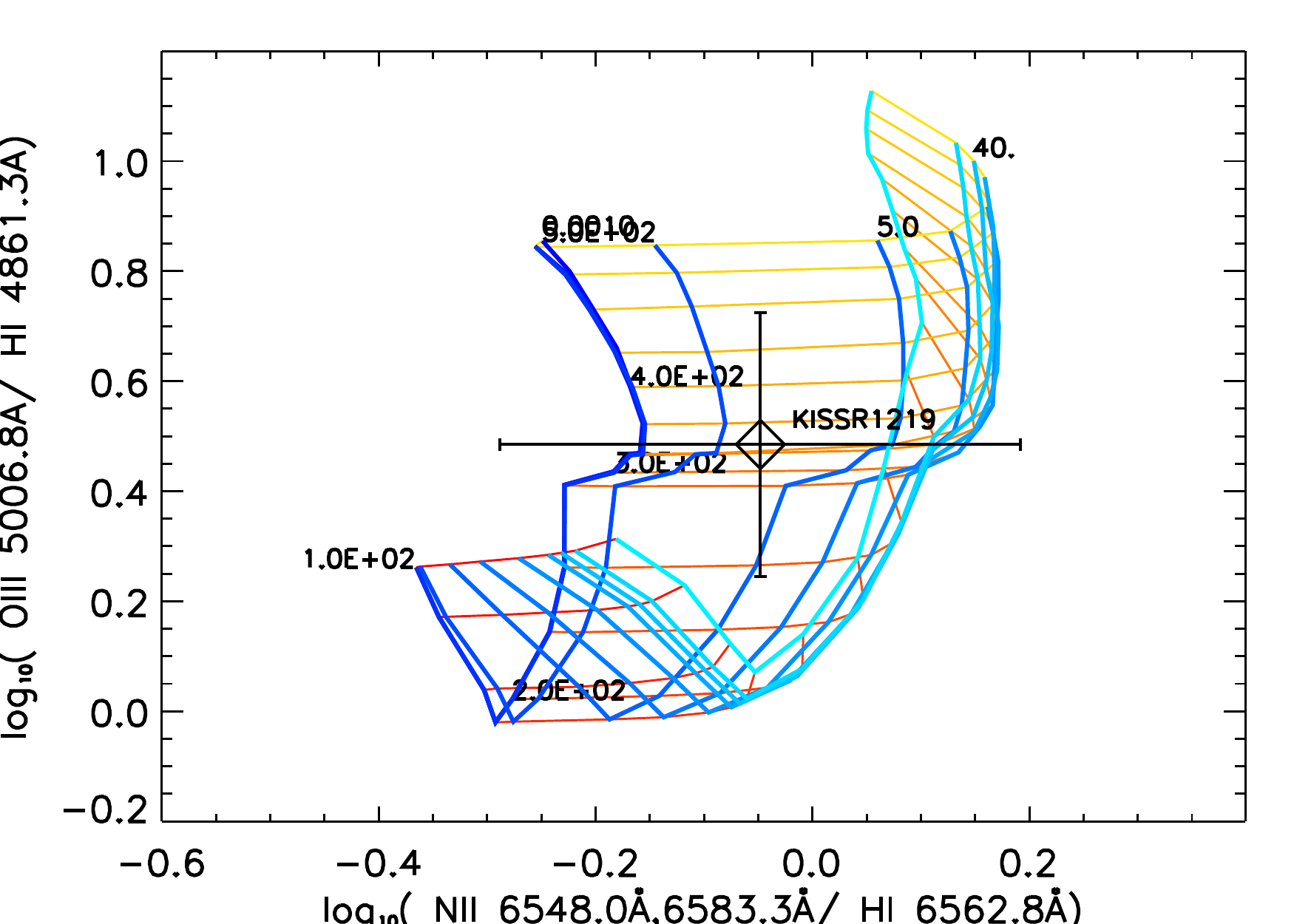}
\caption{[N {\sc ii}] $\lambda$6583/H$\alpha$ vs [OIII] $\lambda$5007/ H$\beta$.  Details same as in Figure ~\ref{fig:sii_ha_oiii_hb}.} 
\label{fig:nii_ha_oiii_hb}
\end{figure*}

\subsection{Star-formation Rate}
If we assume that all of the radio luminosity is attributable to star formation, a star formation rate (SFR) can be derived using the relation in \citet{Condon92}. SFR (for stellar masses $\ge$~5~M$_\sun$) was $\sim4.8$~M$_\sun$~yr$^{-1}$ for the 1.5~GHz VLA emission assuming a spectral index of $-0.8$. The star-formation rate can also be estimated using the optical emission lines \citep{Kennicutt98}. The SFR calculated from all the three components of the H$\alpha$ line ($=5.0\times10^{40}$~erg~s$^{-1}$) turns out to be $\sim0.40$~M$_{\sun}$~yr$^{-1}$. This SFR is an order of magnitude lower than the SFR derived from the radio flux density on kpc-scales. Therefore the ``radio excess" implied from these data supports the AGN origin of the radio emission in KISSR\,1219. 

\subsection{Physical Conditions in the NLR}
\label{secnlr}
Basic properties of the line emitting gas can be obtained from the measurements of the emission line fluxes. We have used the combined line flux of all the fitted Gaussian components in the following analysis. The [S {\sc ii}] or [O {\sc ii}] doublet line intensity ratios gives information about the average electron density $n_e$ of the gas.  Using the [SII] $\lambda$$\lambda$6716, 6731 line intensities ($R_{[S {\sc II}]} = 1.162$) and following the standard calibration of \citet{Osterbrock1989}, the electron density of the gas is estimated to be $\sim250$~cm$^{-3}$ at a temperature of $10^4$~K, consistent with the densities observed in Seyfert galaxies\footnote{{Using the extinction-corrected [O {\sc iii}] $\lambda$4363 line flux from the MPA-JHU spectroscopic analysis for SDSS DR13, we estimated the electron temperature to be $\sim1.2\times10^4$~K using the relation, $\frac{F_{\lambda5007}}{F_{\lambda4363}} = \frac{7.73 \exp[{\frac{3.29\times 10^4}{T}}]}{1 + 4.5\times 10^{-4}(n_e/T_e^{1/2})}$. The [O {\sc iii}] $\lambda$4363 line is detected at the $\sim5\sigma$ level in SDSS, but not significantly enough in our stellar-continuum-subtracted pure emission line spectrum.}}
\citep{Ho1996}. The FWHM([O {\sc i}]) $>$ FWHM([S {\sc ii}]), is also consistent with the empirical trend seen in Seyferts \citep{Ho1996}. This trend is an indication of the density stratification in the sense that dense material is closer to the center. Since [S {\sc ii}] probes low density region, the density stratification suggests that this gas is located far from the center.

\subsection{Dust Extinction \& Mass Outflow Rate}
The ratio of Balmer lines $H\alpha/H\beta$ or Balmer decrement, estimates the attenuation of the emission lines due to dust along the line of sight. In low-density regions of AGN, a value of 3.1 is often used \citep{Osterbrock2006}.  Any deviation from this value is a signature of dust extinction. The ratio of $H\alpha$ to $H\beta$ for KISSR\,1219 is 6.25 indicating the presence of dust distributed as a foreground screen. The optical extinction $A_V$ towards this galaxy is obtained to be 2.2. We used the galactic extinction law based on \citet{Cardelli1989}.

The gas mass of the NLR can be expressed as, $M_{gas} = \frac{m_p L_{H{\alpha}}}{n_e j_{H_{\alpha}}(T)},$
where $m_p$ is the mass of the proton, $n_e\sim250$~cm$^{-3}$ is the electron density, $j_{H_{\alpha}}(T) = 3.53\times10^{-25}$~cm$^3$~erg~s$^{-1}$ is the emission coefficient.  The calculated ionized gas mass is $\sim4.7\times 10^5$~M$_{\odot}$. If we assume that this gas is being {pushed out}, the dynamical time scale of the gas flowing out with an average outflow velocity of  $\sim300$~km~s$^{-1}$ to the edge of the NLR (in this case, $>$2.2 kpc which is the size of the SDSS fibre) is $t_{dyn} = r/v = 7$~Myr.  The corresponding mass outflow rate thus is $\dot{M}_{out} = M_{gas}/t_{dyn} = 0.06$~M$_{\odot}$~yr$^{-1}$. The mass outflow rate is an order of magnitude lower than the current SFR in the galaxy. The gas outflow therefore does not seem to have a significant impact on the SFR of the galaxy.

\subsection{Ionization of the NLR gas}
Line ratio diagrams are powerful tools to investigate the physics of emission line gas. Using grids of theoretical models, one can estimate the parameters such as shock velocity, ionization parameter, or chemical abundance.  The standard optical diagnostic diagrams  include [O {\sc iii}] $\lambda$5007/H$\beta$ vs [N {\sc ii}] $\lambda$6583/H$\beta$,  $\lambda$5007/H$\beta$ vs [S {\sc ii}] $\lambda$$\lambda$6716,6731/H$\alpha$, or [O {\sc i}] $\lambda$6000/{H$\beta$} $\lambda$5007 vs [O {\sc iii}] $\lambda$5007/H$\beta$ \citep{Veilleux1987}. MAPPINGS III shock and photoionization modelling code has been used to predict the line ratios \citep{Dopita1996,Allen2008}. We use the IDL Tool for Emission-line Ratio Analysis \citep[ITERA;][]{Groves2010} for generating line ratio diagrams. These line ratios were studied for several models including dust-free AGN photoionization, dusty AGN photoionization, shock-only, precursor-only, and shock+precursor models \citep{Groves2004,Allen2008}. 

Shocks can be a powerful source of generating ionizing photons. The kinetic energy of the shock compresses and heats the ambient gas, producing a significant fraction of the EUV and X-ray photon field via free-free emission. While slow moving gas just heats up the gas, fast radiative shocks can generate a stronger radiation field. The photoionization front of this strong radiation field increases rapidly and exceeds that of the shock front, thereby separating itself from the shock front. This photoionization front forms the ``precursor'' H{\sc ii} region in the upstream gas. The emission in this precursor region dominates the emission in the shock-only region. A combination of the radiative shocks and the precursor region results in a stronger radiation field and a harder spectrum. Figures~\ref{fig:sii_ha_oiii_hb}, \ref{fig:oi_ha_oiii_ha}, \ref{fig:nii_ha_oiii_hb} show the standard optical line ratio diagrams. The line ratios of KISSR\,1219 lie in the general area expected for Seyfert galaxies in the BPT diagram. The observed line ratios are reasonably well explained by both dusty AGN {photoionization} as well as fast radiative shocks ($v_s\sim300$~km~s$^{-1}$, see Figure~$5-7$) with a precursor, likely driven by the AGN jet. 

We note that in Figure ~\ref{fig:nii_ha_oiii_hb}, the observed [N {\sc ii}] line ratio appears to not fit the dusty AGN photoionization model. This however, can be attributed to the higher abundance of N \citep[at least a factor of 2; ][]{Osterbrock1989} and the observed [N {\sc ii}] line ratio can be brought into agreement with the dusty AGN photoionization model by increasing the N abundance. In the shock+precursor model, however, the strength of the ionizing field does not depend significantly on the atomic abundance. {Therefore, both the models are viable for explaining the line ratios in KISSR\,1219.}

\section{Discussion}
\label{secdisc}
The parsec-scale ``core'' has a steep {$1.5-5$~GHz} spectral index in KISSR\,1219. \citet{Kharb15b} had detected a steep spectrum core also in the DPAGN KISSR\,1494; the core in KISSR\,1494 was suggested to be the unresolved base of a coronal wind, rather than a relativistic jet. However, it is difficult to suggest the same in KISSR\,1219 because of two additional jet components that are observed at the same P.A. as the large-scale radio jet. The coronal wind emission is unlikely to be discrete or collimated. The matching P.A.s between parsec-scale and kpc-scale jet emission, also rules out the three blobs {from being} large stellar complexes. The steep spectrum of the VLBA ``core'' is suggestive of contribution from partially-resolved jet emission, in close proximity to the true unresolved base of the radio jet.

Interestingly, the kpc-scale jet in KISSR\,1219 is reminiscent of the one-sided jet observed in the Seyfert galaxy Arp\,102B \citep{Fathi11}, which is a well-known DPAGN \citep{Eracleous03}. The double peaks are however observed in the broad lines in the Seyfert 1 galaxy Arp\,102B unlike the double peaks in the narrow lines in the Seyfert 2 galaxy KISSR\,1219. Akin to KISSR\,1219, the one-sided radio jet in Arp\,102B is close to one of the spiral arms in this galaxy (but not exactly spatially coincident). We {speculate} if the similarities in the parsec-scale and kpc-scale radio emission and the emission-line spectra, suggest a common physical origin. One of the most favoured mechanism for double peaked emission lines is that the emission line gas is in a rotating disk. While this is likely to be the case for the broad lines in Arp\,102B \citep{Chen89}, having a disky NLR sounds less probable, although it has been suggested in the literature \citep{Mulchaey96}. 

We rather suggest that both KISSR\,1219 and Arp\,102B have relatively powerful radio outflows which are pushing the surrounding emission line clouds and affecting their kinematics; the jet one-sidedness could be a result of Doppler boosting and dimming effects {\citep[as was indeed suggested for Arp\,102B by][]{Fathi11}.}

While the one-sided radio jet detected by the VLA and the VLBA in KISSR\,1219 appears to be very faint, its one-sidedness can be explained by Doppler boosting effects with reasonable jet speeds {($\gtrsim0.55c~- \gtrsim0.25c$} going from parsec to kpc-scales) and orientation angle ($\gtrsim50\degr$). The presence of such powerful outflows should cause the emission line peaks to split as the jets push the gas clouds in opposite directions. A clear blue-shifted broad component is observed in the [O {\sc iii}] emission line (see Figures~\ref{figspectra1} and \ref{figspectra2}), which is indicative of an outflow. Its velocity is $\sim$350~km~s$^{-1}$. The [O {\sc i}] emission lines are also indicative of shock-heated gas. We find that the double peaks are typically separated by about $\sim$200$-$300~km~s$^{-1}$ in different narrow lines. The outflow speeds of the emission lines are therefore lower by factors of several hundred, compared to the radio jet speed. This could indicate that most of the emission line gas is being pushed away in a lateral direction, rather than from the front of the jet, which presumably also has a smaller working surface. The detailed line ratio study using the MAPPINGS III code  suggests that a shock+precursor model can explain the line ionization data well; however the dusty AGN photoionization model cannot be ruled out. If the dusty AGN photoionization model is correct, then the emission line gas could be in broader bipolar outflow surrounding the jet.

We suggest that relatively powerful radio outflows which are pushing the surrounding emission line clouds and affecting their kinematics could also be present in other Seyfert galaxies showing double peaked emission lines and one-sided radio jets, like Arp\,102B.

\section{Summary and Conclusions}
We have carried out VLA observations at 1.5~GHz and phase-referenced VLBI observations at 1.5 and 5~GHz of the double emission line peaked Seyfert galaxy, KISSR\,1219. The primary findings are:
\begin{enumerate}
\item A one-sided radio jet is observed on kpc-scales with the VLA and on parsec-scales with the VLBA at 1.5~GHz. Doppler boosting/dimming with jet speeds of {$\gtrsim0.55c$ to $\gtrsim0.25c$}, going from parsec- to kpc-scales, at a jet inclination $\gtrsim50\degr$ can explain the jet one-sidedness. The time-averaged kinetic power of the radio outflow in KISSR\,1219 is $\approx2.5\times10^{41}$~erg~s$^{-1}$. 
{While the electron densities required to free-free absorb the radio counter-jet emission exist in the NLR clouds or HII regions, their much smaller volume filling factors do not favour them as candidates for absorbers.}
\item The star-formation rate derived from the H$\alpha$ line ($\sim0.40$~M$_{\sun}$~yr$^{-1}$) is an order of magnitude lower than the star-formation rate derived from the radio flux density on kpc-scales ($\sim4.8$~M$_{\sun}$~yr$^{-1}$). The ``radio excess" implied by these data therefore supports an AGN origin for the radio emission in KISSR\,1219. 
\item The weak core-jet structure that is observed with the VLBA at 1.5~GHz, is not detected at 5~GHz. The corresponding $1.5-5$~GHz spectral indices are expected to be steeper than $-1.2, -1.1$ and $-1.0$ for the core, J1 and J2, respectively, with an error of $\sim20\%$. 
{The steep spectrum of the VLBA core is suggestive of contribution from partially-resolved jet emission in close proximity to the true unresolved base of the radio jet.} The relatively high brightness temperatures of the core, J1 and J2 ($\sim6\times10^6$~K, $\sim3\times10^6$~K, and $\sim4\times10^6$~K, respectively), are consistent with non-thermal AGN related emission. The absence of dual parsec-scale radio cores puts the binary BH picture in doubt for the case of KISSR\,1219.
\item The mass of the black hole in KISSR\,1219 is $2.1\pm1.5\times10^7~\mathrm{M_{\sun}}$ using the $\mathrm{M_{BH}}-\sigma_\star$ relation. It is accreting at an Eddington rate of $\sim0.02$, typical of low luminosity Seyfert galaxies. Using the emission lines, we also derive a mass accretion rate of $\sim0.01$~M$_{\odot}$~yr$^{-1}$ in KISSR\,1219, for a mass-to-energy conversion efficiency factor of 0.1. 
\item A blue-shifted broad emission line component in [O {\sc iii}] is indicative of an outflow in the emission line gas at a velocity of $\sim350$~km~s$^{-1}$, while the [O {\sc i}] doublet lines suggest the presence of shock-heated gas. A detailed line ratio study using the MAPPINGS III code further suggests that a shock+precursor model can explain the line ionization data well. Overall, our data suggest that the radio outflow is pushing the emission line clouds, both ahead of the jet and in a lateral direction, giving rise to the double peak emission line spectra in this Seyfert galaxy. The emission line ratios in KISSR\,1219 are indeed consistent with ionization via shocks with a precursor. Future Integral Field Unit observations at optical wavelengths, which can provide spatial information on the emission line gas, will be crucial to test the above suggestions. 
\end{enumerate}

\acknowledgments
{We thank the referee for making insightful suggestions that have improved this manuscript. 
S.S acknowledges research funding support from Chinese Postdoctoral Science Foundation (grant number 2016M590013).}
The National Radio Astronomy Observatory is a facility of the National Science Foundation operated under cooperative agreement by Associated Universities, Inc. This research has made use of the NASA/IPAC Extragalactic Database (NED) which is operated by the Jet Propulsion Laboratory, California Institute of Technology, under contract with the National Aeronautics and Space Administration. Funding for the SDSS and SDSS-II has been provided by the Alfred P. Sloan Foundation, the Participating Institutions, the National Science Foundation, the U.S. Department of Energy, the National Aeronautics and Space Administration, the Japanese Monbukagakusho, the Max Planck Society, and the Higher Education Funding Council for England. 

{\it Facilities:} VLBA, VLA, Sloan.

\bibliographystyle{apj}
\bibliography{ms}

\begin{table*}
\caption{Fitted Line Parameters for KISSR\,1219}
\begin{center}
\begin{tabular}{lclcllc}
\hline\hline
{Line} & {$\lambda_{0}$}  & {$\lambda_{c}\pm$error} & {$\Delta\lambda\pm$error} & {$f_{p}\pm$error} & {$F\pm$error}& {$L\pm$error}\\
{(1)}     & {(2)}    & {(3)}                                  & {(4)} & {(5)} & {(6)} & {(7)}\\ \hline
$[\mathrm {SII}]$ &6718.3 &   6717.05 $\pm$   1.62 &   2.39 $\pm$   0.23 &  31.58 $\pm$   4.09 & 189.22 $\pm$  30.53 &   0.52 $\pm$   0.08\\
& &   6721.71 $\pm$   2.03 &   1.96 $\pm$   0.36 &  18.24 $\pm$   2.58 &  89.83 $\pm$  20.80 &   0.24 $\pm$   0.06\\
& &   6715.85 $\pm$   1.78 &   7.22 $\pm$   1.11 &   8.35 $\pm$   1.83 & 151.03 $\pm$  40.45 &   0.41 $\pm$   0.11\\
$[\mathrm {SII}]$& 6732.7&   6731.05 $\pm$   1.62 &   2.40 $\pm$   0.23 &  27.11 $\pm$   3.79 & 162.82 $\pm$  27.62 &   0.44 $\pm$   0.08\\
& &   6735.72 $\pm$   2.03 &   1.97 $\pm$   0.36 &  15.66 $\pm$   2.39 &  77.32 $\pm$  18.45 &   0.21 $\pm$   0.05\\
& &   6729.85 $\pm$   1.78 &   7.23 $\pm$   1.11 &   7.17 $\pm$   1.70 & 129.90 $\pm$  36.73 &   0.35 $\pm$   0.10\\
 $[\mathrm {NII}]$& 6549.9&   6548.60 $\pm$   0.10 &   2.32 $\pm$   0.22 &  41.92 $\pm$   7.98 & 243.41 $\pm$  51.95 &   0.66 $\pm$   0.14\\
& &   6554.00 $\pm$   0.09 &   1.90 $\pm$   0.45 &  24.21 $\pm$   0.41 & 115.25 $\pm$  27.65 &   0.31 $\pm$   0.08\\
& &   6543.84 $\pm$   0.53 &   7.03 $\pm$   1.08 &  11.08 $\pm$   3.58 & 195.19 $\pm$  69.84 &   0.53 $\pm$   0.19\\
 $[\mathrm {NII}]$& 6585.3&   6583.61 $\pm$   0.10 &   2.33 $\pm$   0.22 & 123.65 $\pm$  23.55 & 722.81 $\pm$ 154.25 &   1.97 $\pm$   0.42\\
& &   6589.03 $\pm$   0.11 &   1.91 $\pm$   0.35 &  71.43 $\pm$  14.87 & 342.46 $\pm$  94.98 &   0.93 $\pm$   0.26\\
& &   6578.82 $\pm$   0.60 &   7.07 $\pm$   1.09 &  32.69 $\pm$  10.56 & 578.97 $\pm$ 207.15 &   1.58 $\pm$   0.56\\
 H$\alpha$& 6564.6&   6563.48 $\pm$   0.07 &   2.32 $\pm$   0.22 & 138.71 $\pm$  26.42 & 807.78 $\pm$ 172.39 &   2.20 $\pm$   0.47\\
& &   6567.44 $\pm$   0.11 &   1.90 $\pm$   0.35 &  80.13 $\pm$  16.68 & 382.46 $\pm$ 106.08 &   1.04 $\pm$   0.29\\
& &   6562.99 $\pm$   0.41 &   7.05 $\pm$   1.69 &  36.67 $\pm$  11.85 & 647.88 $\pm$ 260.45 &   1.76 $\pm$   0.71\\
 H$\beta$& 4862.7&   4862.14 $\pm$   0.13 &   1.54 $\pm$   0.15 &  33.36 $\pm$   6.35 & 129.10 $\pm$  27.55 &   0.35 $\pm$   0.08\\
& &   4864.81 $\pm$   0.21 &   1.19 $\pm$   0.22 &  19.27 $\pm$   4.01 &  57.38 $\pm$  15.92 &   0.16 $\pm$   0.04\\
& &   4859.91 $\pm$   0.47 &   1.18 $\pm$   0.40 &   8.82 $\pm$   2.85 &  26.17 $\pm$  12.23 &   0.07 $\pm$   0.03\\
$[\mathrm {OIII}]$& 4960.3 &   4957.04 $\pm$   0.11 &   2.06 $\pm$   0.12 &  26.60 $\pm$   2.21 & 137.46 $\pm$  13.94 &   0.37 $\pm$   0.04\\
& &   4961.03 $\pm$   0.15 &   2.08 $\pm$   0.21 &  15.37 $\pm$   1.15 &  79.97 $\pm$  10.06 &   0.22 $\pm$   0.03\\
$[\mathrm {OIII}]$& 5008.2 &   5004.95 $\pm$   0.15 &   2.09 $\pm$   0.21 &  78.48 $\pm$   6.63 & 410.60 $\pm$  53.95 &   1.12 $\pm$   0.15\\
& &   5008.98 $\pm$   0.20 &   2.10 $\pm$   1.19 &  45.34 $\pm$   3.57 & 238.87 $\pm$ 136.55 &   0.65 $\pm$   0.37\\
$[\mathrm {OIII}]$~o& 5008.2 &   5000.67 $\pm$   0.56 &   7.94 $\pm$   0.30 &  31.54 $\pm$   4.56 & 628.01 $\pm$  93.85 &   1.71 $\pm$   0.26\\
$[\mathrm {OIII}]$~o& 4960.3 &   4953.24 $\pm$   1.19 &   6.51 $\pm$   0.82 &   8.73 $\pm$   1.52 & 142.47 $\pm$  30.61 &   0.39 $\pm$   0.08\\
H$\gamma$& 4341.7&   4341.60 $\pm$   0.28 &   1.28 $\pm$   0.12 &  13.25 $\pm$   2.52 &  42.54 $\pm$   9.08 &   0.12 $\pm$   0.02\\
& &   4344.03 $\pm$   0.47 &   0.93 $\pm$   0.17 &   7.65 $\pm$   1.59 &  17.85 $\pm$   4.95 &   0.05 $\pm$   0.01\\
& &   4337.26 $\pm$   1.45 &   2.27 $\pm$   0.76 &   3.50 $\pm$   1.13 &  19.95 $\pm$   9.28 &   0.05 $\pm$   0.03\\
$[\mathrm {OI}]$& 6302.0&   6298.40 $\pm$   2.79 &   4.20 $\pm$   1.90 &   5.13 $\pm$  20.32 &  54.00 $\pm$ 215.18 &   0.15 $\pm$   0.59\\
& &   6302.84 $\pm$   1.49 &   2.97 $\pm$   2.79 &  11.28 $\pm$   5.93 &  83.99 $\pm$  90.42 &   0.23 $\pm$   0.25\\
$[\mathrm {OI}]$&6365.5 &   6362.38 $\pm$   1.45 &   4.24 $\pm$   2.79 &   1.15 $\pm$   5.93 &  12.24 $\pm$  63.58 &   0.03 $\pm$   0.17\\
& &   6366.87 $\pm$   1.49 &   3.00 $\pm$   1.88 &   5.84 $\pm$   5.55 &  43.94 $\pm$  50.03 &   0.12 $\pm$   0.14\\                                                                                                       
\hline
\end{tabular}
\end{center}
{Column~1: Emission lines that were fitted with Gaussian components. {``o" indicates the outflow components.} Column~2: Rest wavelength in vacuum in $\AA$. Columns~3, 4: Central wavelength and line width in $\AA$ along with respective errors. Column~5: Peak line flux in units of $10^{-17}$~erg~cm$^{-2}$~s$^{-1}~\AA^{-1}$ with error. Column~6: Total line flux in $10^{-17}$~erg~cm$^{-2}$~s$^{-1}~\AA^{-1}$. Column~7: Line luminosity in units of $10^{40}$~erg~s$^{-1}$.}
\label{tabprop}
\end{table*}

\begin{table*}
\caption{Equipartition Estimates}
\begin{center}
\begin{tabular}{ccccccccc}
\hline\hline
{Jet} & {$L_{rad}$}              &{$\phi$} &   {$P_{min}$}          & {$E_{min}$}            & {$B_{min}$}& {$E_{tot}$}               & {$U_{tot}$} & {$t_e$}\\
\hline
VLBA & $1.4\times10^{41}$ & 1.0    & $1.9\times10^{-7}$ & $3.7\times10^{53}$ & 1.4              & $4.6\times10^{53}$ & $4.2\times10^{-7}$ & 0.012  \\
          & $1.4\times10^{41}$ &  0.5   & $2.9\times10^{-7}$ & $2.7\times10^{53}$ & 1.8              & $3.4\times10^{53}$ & $6.2\times10^{-7}$ & 0.009 \\
VLA & $1.0\times10^{41}$ & 1.0    & $6.4\times10^{-11}$ & $6.0\times10^{55}$ & 0.026        & $7.5\times10^{55}$ & $1.4\times10^{-10}$ & 4.75  \\
          & $1.0\times10^{41}$ & 0.5    & $9.6\times10^{-11}$ & $4.5\times10^{55}$ & 0.032        & $5.6\times10^{55}$ & $2.1\times10^{-10}$ & 3.56 \\
\hline
\end{tabular}
\end{center}
{Column~1: Total radio luminosity in erg~s$^{-1}$. Column~2: Plasma filling factor. Column~3: Minimum pressure in dynes~cm$^{-2}$.
Column~4: Minimum energy in ergs. Column~5: Minimum B-field in mG. Column~6: Total energy in particles and fields, $E_{tot}$ ($=1.25\times E_{min}$) in ergs. Column~7: Total energy density, $U_{tot}=E_{tot}(\phi V)^{-1}$ in erg~cm$^{-3}$. {Column~8: Electron lifetimes in Myr. See Section~\ref{seckpcjet} for details.}}
\label{tabequip}
\end{table*}

\end{document}